# Controlled Smooth Edge Formation of Graphene Nanoribbons

Deepika[1], T.J. Dhilip Kumar[2], Nitin K. Goel[3], Rakesh Kumar[1*]

[1] *Department of Physics, Indian Institute of Technology Ropar, Rupnagar 140001, INDIA*
[2] *Department of Chemistry, Indian Institute of Technology Ropar, Rupnagar 140001, INDIA*
[3] *Infinera India Pvt. Ltd., Prestige solitaire Level 4, 6, Brunton Road, Bangalore 560025, INDIA*
[*] *email: rakesh@iitrpr.ac.in*

We report energy estimated to dissociate a C-C bond of a graphene sheet to form nanoribbons of armchair and zigzag configurations using first principles calculations. For the ground state energy calculations, the configurations considered are with spin, and without spin polarization. It is observed that the energy required to dissociate a C-C bond of a graphene sheet to form zigzag configuration is higher than that of armchair configuration for both spin polarized state, as well as non-spin polarized state. Therefore, formation of smooth edged graphene nanoribbons along the crystallographic directions might be engineered by a control over energy.

## INTRODUCTION

Graphene[1], a single layer of carbon atoms arranged in two dimensional honeycomb lattices, has attracted a lot of attention due to its extraordinary electronic properties like room temperature Quantum Hall effect[2], and the highest mobility of the charge carriers in suspended graphene[3], but it is a zero band gap material[1]. Therefore, a band gap like semiconductors is required to be opened for electronic applications. One of the possible ways to open a band gap in graphene is Quantum confinement of the charge carriers, which can be achieved by cutting a graphene sheet into nanoribbons. Theoretical studies of graphene nanoribbons (GNRs) show that in addition to quantum confinement, smoothness of crystallographic edges (armchair or zigzag) plays an important role in deciding its semiconducting or metallic behaviour[4], but still it is an open challenge for experimentalists to develop a simple and reliable technique to make GNRs of smooth edges.

The motivation of the present work is to find an external parameter, which can be used to design an experiment in realizing GNRs of well defined crystallographic edges. For electronic calculations, we have used energy as an external parameter. On the basis of first principles calculations, we report the minimum energy required to dissociate a C-C bond of a graphene sheet along armchair and zigzag directions, respectively to form GNRs of smooth crystallographic edges. To study the edge effects on the bond dissociation energy, we have avoided the contribution due to quantum confinement by considering GNRs of equal length and width. On the basis of our results, we remark that energy might be used as an explicit parameter to realise GNRs with smooth crystallographic edge.

## COMPUTATIONAL DETAILS

We have performed first principles calculations based on Density Functional Theory (DFT)[5] using Vienna *ab initio* Simulation Package (VASP)[6] to study the periodic structure of a graphene sheet, armchair graphene nanoribbons (AGNRs), and zigzag graphene nanoribbons (ZGNRs). We used number of dimer lines ($N_a$) in AGNRs, and number of zigzag chains ($N_z$) in ZGNRs to represent the width[7] of corresponding GNRs as $N_a$-AGNRs, and $N_z$-ZGNRs, respectively. To describe the electronic wave functions of the configurations, Plane wave basis set was applied for valence electron orbitals, Generalized Gradient Approximation (GGA) for electrons exchange-correlations, and Projected-Augmented Wave (PAW) potentials for electron-core interactions. To minimize the interlayer interactions, a vacuum layer of 10 Å was used. To study the electronic properties in reciprocal space, Brillouin zone was divided into k-mesh of 25x25x1 for graphene sheet and 25x1x1 for GNRs using Monkhorst–Pack of k-mesh sampling. Force tolerance of 0.001 eV/ Å was used for each atom of the unit cell to relax the configurations. An energy convergence limit of $10^{-5}$ eV, and kinetic cut off energy of 300 eV was used for graphene sheet and GNRs. The k-mesh and cut off energy provided for relaxation and self-consistent calculations was optimized during the pre-analysis calculations. Gaussian 09 package was used with CCSD(T)/cc-pVTZ as basis set[8] to estimate C-H bond formation energy.

## RESULTS AND DISCUSSION

We have performed ground state energy calculations for the unit cell of graphene sheet, and GNRs of armchair and zigzag edge configurations. The carbon atoms at the edges of GNRs are passivated with Hydrogen atoms to stabilize the dangling bonds. The energy released to form C-H bond at the edge of graphene nanoribbons was approximated using bond dissociation energy of the methane molecule (equation 1). Therefore, actual ground state energy of the GNRs was determined by subtracting the corresponding hydrogenation energy at the edges. Dissociation energy per C-C bond to form GNRs from a graphene sheet has been calculated by equation 2.

$$E_{C-H} = \frac{E_{Carbon} + 4*E_{Hydrogen} - E_{Methane}}{4} \quad (1)$$

$$E^{BDE} = \frac{E_{GNR} + n*E_{C-H} - E_{Graphene\ sheet}}{n} \quad (2)$$

In equation 2, $E^{BDE}$ is the dissociation energy of C-C bond to form GNRs from a graphene sheet, n is the number of dissociated C-C bonds. The hydrogenation energy of 4.172 eV (per C-H bond) has been estimated from the methane molecule including zero point correction. Theoretical studies show that the ground state energy of GNRs depends upon its width as well as its edge configurations[9]. Therefore, to compare the edge formation energy of armchair and zigzag configurations, we considered AGNRs and ZGNRs of nearly the same width; thus avoiding the contribution due to quantum confinement.

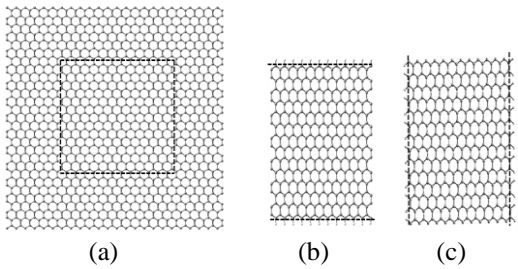

**FIGURE 1.** (Color online), (a) Periodic unit cell of graphene sheet is shown by enclosed dotted lines. Periodic unit cell of Hydrogen terminated edges for 14-ZGNR and 24-AGNR are shown in (b) and (c), respectively. Carbon atoms are shown in gray color and hydrogen atoms in white. Dotted lines represent the periodicity direction.

Here, we discuss the results of ZGNRs and AGNRs for a typical width of 2.84 nm, and 2.83 nm, respectively. Width of 2.84 nm corresponds to 14-ZGNRs (Figure 1(b)), and 2.83 nm to 24-AGNRs (Figure 1(c)), respectively. The parent graphene sheet corresponds to a dimension of 2.84 nm x 2.83 nm (Figure 1(a)). The ground state energy for the unit cell of graphene sheet, and the corresponding GNRs of armchair and zigzag configuration is provided in Table 1.

**TABLE 1.** Ground state energy for periodic unit cell of graphene sheet, Hydrogen terminated armchair and zigzag graphene nanoribbons.

| Periodic Configuration | Width (nm) | Ground state energy (eV) |
|---|---|---|
| (14x24) graphene sheet (spin polarized) | 2.84x2.83 | -3118.710 |
| 24-AGNR (length 2.84 nm) (spin polarized) | 2.83 | -3211.168 |
| 14-ZGNR (length 2.83 nm) (Spin polarized) | 2.84 | -3194.528 |
| (14x24) graphene sheet (non spin polarized) | 2.84x2.83 | -3118.710 |
| 24-AGNR (length 2.84 nm) (non spin polarized) | 2.83 | -3211.168 |
| 14-ZGNR (length 2.83 nm) (non spin polarized) | 2.84 | -3193.783 |

To find energy required for the formation of GNRs of the corresponding configurations from a graphene sheet, the actual ground state energy of the GNRs is subtracted from the ground state energy of the graphene sheet. Hence, total energy involved in the formation of the corresponding GNRs is determined by dissociation of associated C-C bonds in the graphene sheet. The dissociation energy per C-C bond for the formation of armchair and zigzag configurations from a graphene sheet has been estimated to be 0.870 eV and 1.044 eV, respectively for non-spin polarized states. The dissociation energy per C-C bond for spin polarized states to form armchair configuration remains the same as that of non-spin polarized state, but for zigzag configuration it decreased to 1.013 eV due to magnetic contributions of edge states[10] (Table 2). We estimated dissociation energy of C-C bonds of a graphene sheet corresponding to different dimensions to form AGNRs and ZGNRs of nearly the same width. We observe nearly the same results, and hence similar patterns as that of typical graphene sheet considered.

**TABLE 2.** Bond formation/ dissociation energy.

| Bond (formed/ dissociated) | Energy (eV) |
|---|---|
| C-H bond formation energy | -4.172 |
| C-C bond dissociation energy to form AGNR (spin polarized) | 0.870 |
| C-C bond dissociation energy to form ZGNR (spin polarized) | 1.013 |
| C-C bond dissociation energy to form AGNR (non-spin polarized) | 0.870 |
| C-C bond dissociation energy to form ZGNR (non-spin polarized) | 1.044 |

From the above results, we remark two things; (a) for both spin and non-spin polarized state, energy required to form a zigzag configuration from a graphene sheet is higher than that of an armchair configuration, (b) the difference of the dissociation energy per C-C bond to form AGNRs and ZGNRs configurations from a graphene sheet is considerable enough to experimentally manoeuvre the formation atomically smooth edges of AGNRs and ZGNRs. From these observations, it can be concluded that energy can be used as a controlling parameter in the formation of smooth edged GNRs.

## ACKNOWLEDGMENT

Authors would like to thank whole team of Param Yuva II, CDAC-Pune for providing supercomputing facility, and IIT Ropar for supporting the research work.